\documentclass[letterpaper, 10 pt, conference]{ieeeconf}  

\IEEEoverridecommandlockouts                              

\usepackage{graphicx} 
\usepackage{cite}
\usepackage{color}
\usepackage{url}
\usepackage{balance}
\newcommand{\etal}{\textit{et al.~}}

\title{\LARGE \bf
1D CNN Architectures for Music Genre Classification
}

\author{Safaa Allamy and Alessandro L. Koerich
\thanks{Safaa Allamy and Alessandro L. Koerich are with \'{E}cole de Technologie Sup\'{e}rieure, Universit\'{e} du Qu\'{e}bec, Montr\'{e}al, QC, H3C 1K3, Canada
        {\tt\scriptsize allamy.safaa@gmail.com,  alessandro.koerich@etsmtl.ca}}%
}

\begin{document}

\maketitle
\thispagestyle{empty}
\pagestyle{empty}

\begin{abstract}
This paper proposes a 1D residual convolutional neural network (CNN) architecture for music genre classification and compares it with other recent 1D CNN architectures. The 1D CNNs learn a representation and a discriminant directly from the raw audio signal. Several convolutional layers capture the time-frequency characteristics of the audio signal and learn various filters relevant to the music genre recognition task. The proposed approach splits the audio signal into overlapped segments using a sliding window to comply with the fixed-length input constraint of the 1D CNNs. As a result, music genre classification can be carried out on a single audio segment or on the aggregation of the predictions on several audio segments, which improves the final accuracy. The performance of the proposed 1D residual CNN is assessed on a public dataset of 1,000 audio clips. The experimental results have shown that it achieves 80.93\% of mean accuracy in classifying music genres and outperforms other 1D CNN architectures.
\end{abstract}

\section{INTRODUCTION}
In the last decade, music has become more and more accessible to everyone across the globe. The exponential growth of musical data has prompted the development of new tools that are more personalized than radio broadcasts, for example. In addition to allowing consumers to listen to their favorite songs, these tools also allow them to discover music from different backgrounds. To such an aim, several approaches to music information retrieval (MIR) have been developed~\cite{Lidy2010music}. MIR focuses on the research and development of computational systems to advance the access, organization, and understanding of music information.

Music genre classification is one of the most popular research topics in MIR, which consists of automating the categorization of music items into pre-defined genres such as rock, pop, jazz, classical, etc. For instance, this is of great importance for music streaming and music shopping services because the recognition of musical genres allows these digital platforms to have a greater analytical capacity and personalize their services to their customers. Given its importance in MIR, classifying musical genres has been the subject of several studies and research in the last two decades~\cite{Tzanetakis2002,costa2004automatic,Turnbull2005,koerich2005combination,Scaringella2006,silla2008machine,Lidy2010music,Dieleman2014,Choi2017,Costa2017,Pons2018,Lee2018,Kim2019,Koerich2020}

However, automatic classification of musical genres is a non-trivial task, and it is somewhat subjective because humans determine musical genres with a high degree of subjectivity. That makes their definition vague because it is not easy to adequately describe sounds using words. Furthermore, it is also possible that individuals disagree on the musical genre of the same song. Therefore, classifying music genres is at first glance a difficult task, even for humans. Also, it is not easy to determine which features to use to accomplish this task. 

Recent studies have relied on CNNs for music genre classification~\cite{Dieleman2014,Choi2017,Costa2017,Pons2018,Lee2018,Kim2019,Koerich2020}. Most of the proposed approaches employ 2D CNNs, which use spectrograms as input. A spectrogram is a visual representation of the spectrum of frequencies of the music signal as it varies with time. The architecture of 2D CNNs used with spectrograms is similar to those used in object recognition and in many other computer vision tasks that process images or videos. However, in recent years, several researchers have proposed 1D CNNs, which use the raw audio signal as input. Different 1D CNN architectures have been proposed for audio tagging~\cite{Kim2019}, music genre classification~\cite{Dieleman2014,Choi2017,Pons2018,Lee2018,Kim2019,Koerich2020}, environmental sound classification~\cite{abdoli2019}.

This paper proposes a 1D residual CNN architecture for music genre classification and compares it with other recent 1D CNN architectures. The main contributions of this paper are: (i) a novel 1D residual CNN architecture that outperforms several 1D CNN architectures; (ii) a comparison of five different 1D CNN architectures to deal with music genre classification on a benchmarking dataset; (iii) a comparison of the performance of 1D CNN architectures with and without data augmentation.

This paper is organized as follows. Section~\ref{sec:rel} presents several recent approaches based on 1D and 2D CNNs that deal with music genre classification. Section~\ref{sec:propo} presents the proposed approach for music genre classification and the proposed 1D residual CNN architecture. Section~\ref{sec:1DCNNs} describes several 1D CNN architectures for audio classification. Section~\ref{sec:exp} presents a general description of the dataset and the experiments carried out to evaluate the proposed 1D residual CNN and the comparison with other 1D CNN architectures. In the last section, we present our conclusions and the perspectives of further work.

\section{RELATED WORK}
\label{sec:rel}
The classification of musical genres has been the subject of several studies in the last two decades~\cite{Tzanetakis2002,costa2004automatic,Turnbull2005,koerich2005combination,Scaringella2006,silla2008machine,Lidy2010music,costa2013music,Costa2017,Kim2018,Koerich2020}. Early studies have employed content-based shallow approaches that extract acoustic features from the music audio to train classification models such as $k$ nearest neighbors, support vector machines, decision trees, etc~\cite{Tzanetakis2002,costa2004automatic,Turnbull2005,koerich2005combination,Lidy2010music,costa2013music}. Different categories of acoustic features such as beat-related, timbral texture, pitch-related, etc., can be extracted directly from the raw audio signal~\cite{Tzanetakis2002,costa2004automatic,koerich2005combination,Scaringella2006,silla2008machine}. The music audio can also be transformed into a spectrogram, a visual representation of the spectrum of frequencies of the music signal as it varies with time, and visual features can be extracted from such spectrograms~\cite{costa2012music,koerich2013improving}. There are several types of spectrograms such as Mel-frequency cepstral coefficient (MFCC), short-time Fourier transform (STFT), discrete wavelet transform (DWT), cross recurrence plot (CRP), among others~\cite{esmaeilpour2020robust,esmaeilpour2021sound}.

In recent years, CNNs have had a significant impact on many audio and music processing tasks~\cite{Dieleman2014,Sainath2015,Collobert2016,Zhu2016,Thickstun2017,Costa2017,Choi2017,Pons2018,Kim2018,Lee2018,Kim2019,abdoli2019, Koerich2020}. For instance, several researchers have employed CNNs for music genre classification~\cite{Dieleman2014,Costa2017,Choi2017,Pons2018,Koerich2020}. However, one of the main bottlenecks for using CNNs in music genre classification is the amount of data required to train such networks properly since they usually have thousands of parameters to adjust. Two main approaches have been used to circumvent this problem: fine-tuning CNNs pre-trained on large datasets even if these datasets lay in other domains; use data augmentation procedures to increase the number of samples artificially by introducing small modifications in the original samples. The first approach is usually used with pre-trained 2D CNNs borrowed from the computer vision domain. CNNs pre-trained in datasets such as ImageNet have their last convolutional layers and fully connected layers fine-tuned on the target datasets. On the other hand, pre-trained 1D CNNs are not very commonly available, and fine-tuning them does not usually yield good results as their 2D counterparts. 

Choi~\etal\cite{Choi2017} introduced a 2D convolutional recurrent neural network (CRNN) for music tagging. The proposed architecture takes advantage of CNNs for local feature extraction and recurrent neural networks for temporal summarization of the extracted features. The CRNN uses log-amplitude Mel-spectrograms as input and a 2-layer RNN with gated recurrent units (GRU) to summarize temporal patterns on the top of 2D 4-layer CNNs. They used the Million Song Dataset (MSD) with last.fm tags. The CRNN was trained to predict the top-50 tag, including genres, moods, instruments, and eras. It achieved an area under the curve (AUC) between 0.84 and 0.86.
Costa~\etal\cite{Costa2017} used a 2D CNN architecture that has achieved high accuracy on different pattern recognition tasks. Such an architecture uses CLs with 5$\times$5 kernels and 64 filters, followed by MPLs and an FCL at the end. The proposed 2D CNN achieved 92\% and 85.9\% accuracy on LMD~\cite{silla2008latin} and ISMIR2004 datasets, respectively. Song~\etal\cite{Song2017} used transfer learning to pre-train a five-layer recurrent
neural network (RNN) with a gated recurrent unit (GRU) on the Magnatagatune dataset. Experimental results show the transfer learning way can achieve a high classification accuracy on GTzan dataset.
Oramas~\etal\cite{Oramas2017} presented a new dataset for musical categorization encompassing audio, text, and images. Additionally, they proposed an approach for multi-label genre classification based on the combination of feature embeddings. They evaluated three 2D CNN architectures with different convolution
filter sizes and a different number of filters per convolutional layer and target layer. They used log amplitude constant-Q transform (CQT) spectrograms as input of the CNNs. The best CNN architecture achieved an AUC of 0.888 on their dataset.

Due to the effectiveness of 2D CNNs in recognizing musical genres, several researchers have focused on exploiting 1D CNN architectures. However, in this case, one of the main bottlenecks is the lack of robust pre-trained models. Furthermore, 1D CNN models seem to be more challenging to fine-tune than their 2D counterparts.  

One of the first 1D CNN architectures for music tagging was proposed by Dieleman and Schrauwen~\cite{Dieleman2014}. They investigated feature learning directly to raw audio signals. They trained both 1D and 2D CNNs and compared their performance on an automatic tagging task. The experimental results showed that 1D CNNs do not outperform 2D CNNs. However, 1D CNNs can discover frequency decompositions from audio waveforms, as well as phase-and translation-invariant feature representations.
Kim~\etal\cite{Kim2018} proposed a CNN architecture that learns representations from tiny grains of waveforms. The experiment results showed that such an architecture improves music auto-tagging accuracy and provides results comparable to previous studies on the same subject using the MagnaTagATune (MTAT) and the million songs (MSD) datasets.
Pons~\etal\cite{Pons2018} proposed an end-to-end 1D CNN architecture for music audio tagging. The study also compared the proposed 1D CNN architecture to a 2D CNN showing that the performance of waveform-based models outperforms spectrogram-based models in the case of large-scale data.

\section{PROPOSED APPROACH}
\label{sec:propo}
The proposed approach aims to handle music tracks of variable lengths. One of the challenges of modeling variable-length signals with CNN architectures is that we must have fixed length inputs. One way to circumvent this constraint is to split the audio signal into several fixed-length segments using a sliding window of appropriate width~\cite{abdoli2019}. Therefore, we adopt a sliding window of fixed width to conditionate the audio signal to the input layer of the 1D CNN.

\begin{figure}[htpb!]
  \centering
  \includegraphics[width=0.48\textwidth]{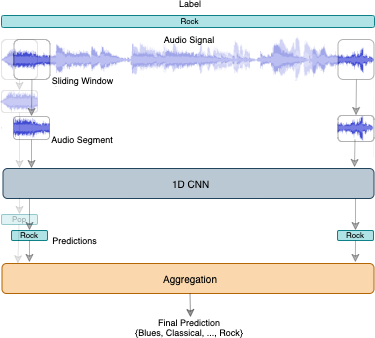}
  \caption{Splitting the input audio signal into several segments with an sliding window with 75\% of overlapping. Adapted from~\cite{Koerich2020}.}
  \label{fig:framing}
\end{figure}

The width of the sliding window depends mainly on the signal sampling rate, and successive audio frames may also have a certain percentage of overlapping. Such overlapping aims to maximize the use of information and augment training data as some audio signal parts are reused. Fig.~\ref{fig:framing} illustrates an overview of the proposed approach.

The audio signal has a single label shared with all audio segments during the training of the 1D CNN. Once the CNN is trained, it can be used to predict the music genre of a new music piece. For prediction, the same sliding window, with or without overlapping, is used to segment the audio signal into audio segments. Thus, the prediction can be carried out on a single audio segment. However, the aggregation of the predictions on several audio segments improves the reliability of the predictions. This strategy also allows predicting the music genre on audio streams.

For all 1D CNN architectures, we consider audio samples at 22,050 Hz and audio segments of 5 seconds. Therefore, the input of the 1D CNNs consists of a 110,250-dimensional array. Such a sampling rate and segment length provide the best trade-off between accuracy and input dimensionality~\cite{Koerich2020}, which affects the number of parameters of the 1D CNNs.

\subsection{1D ResNet CNN}
\begin{figure}[htpb!]
\centering
\includegraphics[scale=0.47]{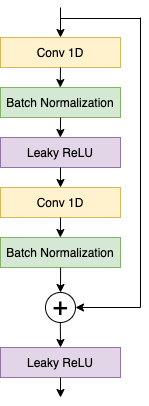}
\caption{The architecture of the one-dimensional residual blocks (Res1D).}
\label{fig:resblock}
\end{figure}

The proposed residual 1D CNN architecture is based on the sample-level CNN proposed by Kim~\etal\cite{Kim2018}, but residual blocks replace most convolutional layers (CLs). The use of residual blocks with skip connections ease training deep networks avoiding degradation and vanishing gradient problems~\cite{He2016}. The architecture of the residual block is shown in Fig.~\ref{fig:resblock}, and it consists of two CLs with filter size three and stride one, two batch normalization (BN) layers, and an identity shortcut. LeakyReLU is used as the activation function. Such a 1D ResNet CNN uses small kernels of size three and stride one.

\begin{table}[htpb!]
\renewcommand{\arraystretch}{1.1}
\caption{The proposed 1D ResNet CNN architecture.}
\label{tab:resnet}
	\centering
		\begin{tabular}{|c|c|c|c|c|c|c|}
			\hline
			 & \bf \# & \bf Kernel & \bf Pool &  & \bf Output \\
			 \bf Layer & \bf Filters  & \bf Size & \bf Size & \bf Stride & \bf Shape\\
			\hline
			Input & - & - & - & - & 110,250\\
			\hline
			Conv1D & 128 & 3 & - & 3 &128$\times$36,750\\
			\hline
			Res1D & 128 & 3 & - & 1 & 128$\times$36,750\\
            \hline
			MaxPool & - & - & 3 & 3 & 128$\times$12,250\\
            \hline
            Res1D & 128 & 3 & - & 1 & 128$\times$12,250\\
            \hline
			MaxPool & - & - & 3 & 3 & 128$\times$4,083\\
            \hline
            Res1D & 256 & 3 & - & 1 & 256$\times$4,083\\
            \hline
			MaxPool & - & - & 3 & 3 & 256$\times$1,361\\
            \hline
            Res1D & 256 & 3 & - & 1 & 256$\times$1,361\\
            \hline
			MaxPool & - & - & 3 & 3 & 256$\times$453\\
            \hline
            Res1D & 256 & 3 & - & 1 & 256$\times$453\\
            \hline
            MaxPool & - & - & 3 & 3 & 256$\times$151\\
            \hline
            Res1D & 256 & 3 & - & 1 & 256$\times$151\\
            \hline
            MaxPool & - & - & 3 & 3 & 256$\times$50\\
            \hline
            Res1D & 256 & 3 & - & 1 & 256$\times$50\\
            \hline
            MaxPool & - & - & 3 & 3 & 256$\times$16\\
            \hline
            Res1D & 256 & 3 & - & 1 & 256$\times$16\\
            \hline
            MaxPool & - & - & 3 & 3 & 256$\times$5\\
            \hline
            Res1D & 512 & 3 & - & 1 & 512$\times$5\\
            \hline
            MaxPool & - & - & 3 & 3 & 512$\times$1\\
            \hline
            Conv1D & 512 & 1 & - & 1& 512$\times$1\\
            \hline
            Output & - & - & - & - & 10\\
            \hline
            \multicolumn{6}{l}{\# Trainable parameters: 4,086,794.}
		\end{tabular}
\end{table}

\section{1D CNN Architectures}
\label{sec:1DCNNs}
Several 1D CNN architectures have been proposed to deal directly with the audio waveforms. In this paper, we have considered five existing 1D CNN architectures\footnote{All 1D CNN architectures describe in this paper are available at \url{https://github.com/alekoe/1DCNNs}} that have been used in audio related tasks.

Kim~\etal\cite{Kim2018} proposed an end-to-end 1D approach to automatic music tagging called sample-level CNN. This approach takes audio waveforms as input. The architecture of this 1D CNN is presented in Table~\ref{tab:samplecnn}, and it is made up of 11 CLs, most of them with filters of size three and stride three, except the last CL, which has filters and stride of size one. A max-pooling layer (MPL) is added after CLs to reduce the dimensionality of the output feature maps. The output layer has ten units and a sigmoid activation function, followed by a dropout of 0.5. All other layers use the ReLU activation function and BN after each CL.

\begin{table}[htpb!]
\renewcommand{\arraystretch}{1.1}
\caption{Sample-level 1D CNN architecture~\cite{Kim2018}.}
\label{tab:samplecnn}
	\centering
		\begin{tabular}{|c|c|c|c|c|c|c|c|}
			\hline
			 & \bf \# & \bf Kernel & \bf Pool &  & \bf Output \\
			 \bf Layer & \bf Filters  & \bf Size & \bf Size & \bf Stride & \bf Shape\\
			\hline
			Input & - & - & - & - & 110,250\\
			\hline
			Conv1D & 128 & 3 & -& 3 & 128$\times$36,750\\
			\hline
			Conv1D & 128 & 3 &-& 1 & 128$\times$36,750\\
            \hline
			MaxPool & - & - & 3 & 3 &128$\times$12,250\\
            \hline
            Conv1D & 256 & 3 & - & 1 &128$\times$12,250\\
            \hline
			MaxPool & - & - & 3 & 3 &128$\times$4,083\\
            \hline
            Conv1D & 256 & 3 & - & 1 &256$\times$4,083\\
            \hline
			MaxPool & - & - & 3 & 3 &256$\times$1,361\\
            \hline
            Conv1D & 256 & 3 & - & 1&256$\times$1,361\\
            \hline
			MaxPool & - & - & 3 & 3 &256$\times$453\\
            \hline
            Conv 1D & 256 & 3 & - & 1 &256$\times$453\\
            \hline
			MaxPool & - & - & 3 & 3 &256$\times$151\\
            \hline
            Conv1D & 256 & 3 & - & 1 &256$\times$151\\
            \hline
			MaxPool & - & - & 3 & 3 &256$\times$50\\
            \hline
            Conv1D & 256 & 3 & - & 1 &256$\times$16\\
            \hline
			MaxPool & - & - & 3 & 3 &256$\times$5\\
            \hline
            Conv1D & 512 & 3 & - & 1&512$\times$5\\
            \hline
			MaxPool & - & - & 3 & 3 &512\\
            \hline
            Conv1D & 512 & 1 & - & 1 &512\\
            \hline
            Output & - & - & - & - & 10 \\
            \hline
            \multicolumn{5}{l}{\# Trainable parameters: 1,848,842}
		\end{tabular}
\end{table}

Similar to the architecture proposed by Kim~\etal\cite{Kim2018}, the architecture proposed by Pons~\etal\cite{Pons2018} also uses small kernel filters. The 1D architecture consists of 7 CLs interchanged with MPLs, as presented in Table~\ref{tab:scalecnn}. The CLs have filters of size three and stride one, except the first CL, which has stride three. BN is employed after each CL. The MPLs have pool size and stride three.

\begin{table}[htpb!]
\renewcommand{\arraystretch}{1.1}
\caption{End-to-end at scale CNN architecture~\cite{Pons2018}.}
\label{tab:scalecnn}
	\centering
		\begin{tabular}{|c|c|c|c|c|c|c|c|}
			\hline
			 & \bf \# & \bf Kernel & \bf Pool &  & \bf Output \\
			 \bf Layer & \bf Filters  & \bf Size & \bf Size & \bf Stride & \bf Shape\\
			\hline
			Input & - & - & - & - & 110,250\\
			\hline
			Conv1D & 64 & 3 & - & 3 & 64$\times$36,750\\
			\hline
			Conv1D & 64 & 3 & - & 1 & 64$\times$36,748\\
            \hline
			MaxPool & - & - & 3 & 3 & 64$\times$12,249\\
            \hline
            Conv1D & 64 & 3 & - & 1 & 64$\times$12,247\\
            \hline
			MaxPool & - & - & 3 & 3 & 64$\times$4,082\\
            \hline
            Conv1D & 128 & 3 & - & 1 & 128$\times$4,080\\
            \hline
			MaxPool & - & - & 3 & 3 &128$\times$1,360\\
            \hline
            Conv1D & 128 & 3 & - & 1 & 128$\times$1,358\\
            \hline
			MaxPool & - & - & 3 & 3 & 128$\times$452\\
            \hline
            Conv 1D & 128 & 3 & - & 1 & 128$\times$450\\
            \hline
			MaxPool & - & - & 3 & 3 & 128$\times$150\\
            \hline
            Conv1D & 256 & 3 & - & 1 &256$\times$148\\
            \hline
			MaxPool & - & - & 3 & 3 &256$\times$49\\
            \hline
            Output & - & - & - & - & 10 \\
            \hline
            \multicolumn{6}{l}{\# Trainable parameters: 373,898}
		\end{tabular}
\end{table}

Dieleman and Schrauwen~\cite{Dieleman2014} also proposed an end-to-end learning approach for automatic music tagging based on a 1D CNN architecture. The proposed 1D CNN has a strided CL followed by two CLs with filters of size eight interchanged with MPLs with pooling size four and two fully connected layers (FCLs) with 100 and 10 units. The last layer uses a sigmoidal activation function, while all other layers use the ReLU activation function. For the strided CL, we used one filter of size 256 and stride 256, which were the parameters that yielded the best performance in~\cite{Dieleman2014}. Such an architecture is summarized in Table~\ref{tab:Dielemancnn}.

\begin{table}[htpb!]
\renewcommand{\arraystretch}{1.1}
\caption{CNN architecture adapted from~\cite{Dieleman2014}.}
\label{tab:Dielemancnn}
	\centering
		\begin{tabular}{|c|c|c|c|c|c|c|c|}
			\hline
			 & \bf \# & \bf Kernel & \bf Pool &  & \bf Output \\
			 \bf Layer & \bf Filters  & \bf Size & \bf Size & \bf Stride & \bf Shape\\
			\hline
			Input & - & - & - & - & 110,250\\
            \hline
			Conv1D & 1 & 256 & - & 256 & 1$\times$430\\
			\hline
			Conv1D & 32 & 8 & - & 2 & 32$\times$212\\
			\hline
			Conv1D & 32 & 8 & - & 1 & 32$\times$205\\
            \hline
			MaxPool & - & - & 4 & 4 & 32$\times$51\\
            \hline
            Conv1D & 32 & 8 & - & 1 & 32$\times$44\\
            \hline
			MaxPool & - & - & 4 & 4 & 32$\times$11\\
            \hline
			FC & - & - & - & - &100\\
            \hline
            Output & - & - & - & - & 10 \\
            \hline
            \multicolumn{6}{l}{\# Trainable parameters: 53,495}
		\end{tabular}
\end{table}

The 1D CNN architectures presented so far have been proposed to deal mainly with music audio. Abdoli~\etal\cite{abdoli2019} proposed an end-to-end 1D CNN architecture for environmental sound classification (ESC) that can process audio signals of varying lengths. The 1D CNN initializes the first CL with a Gammatone filter bank. A gamma filter is a linear filter described by an impulse response that is the product of a gamma distribution and a sinusoidal tone. Such a filter bank models the human auditory filter response in the cochlea. The proposed 1D CNN uses large kernels (512) in the first CL since it is assumed that the first layer should have a more global view of the audio signal. The authors stated that shorter filters do not provide a general view of the signal's spectral contents. The proposed architecture comprises four CLs interlaced with MPLs, followed by two FCLs and an output layer. The output layer uses a sigmoidal activation function, while all other layers use the ReLU activation function. BN is employed after each CL. A summary of the 1D CNN architecture is shown in Table~\ref{tab:Abdolicnn}.

\begin{table}[htpb!]
\renewcommand{\arraystretch}{1.1}
\caption{1D CNN architecture for ESC proposed by Abdoli~\etal\cite{abdoli2019}.}
\label{tab:Abdolicnn}
	\centering
		\begin{tabular}{|c|c|c|c|c|c|c|c|}
			\hline
			 & \bf \# & \bf Kernel & \bf Pool &  & \bf Output \\
			 \bf Layer & \bf Filters  & \bf Size & \bf Size & \bf Stride & \bf Shape\\
			\hline
			Input & - & - & - & - & 110,250\\
			\hline
			Conv1D$^*$ & 64 & 512 & - & 1 & 64$\times$109,739\\
			\hline
			MaxPool & - & - & 8 & 8 & 64$\times$13,717\\
            \hline
            Conv1D & 32 & 32 & - & 2 & 32$\times$6,859\\
            \hline
			MaxPool & - & - & 8 & 8 & 32$\times$857\\
            \hline
            Conv1D & 64 & 16 & - & 2 & 64$\times$429\\
            \hline
            Conv1D & 128 & 8 & - & 2 & 128$\times$215\\
            \hline
            Conv1D & 256 & 4 & - & 2 & 256$\times$108\\
            \hline
			MaxPool & - & - & 4 & 4 & 256$\times$27\\
            \hline
			FC & - & - & - & - &128\\
            \hline
			FC & - & - & - & - &64\\
            \hline
            Output & - & - & - & - & 10 \\
            \hline
            \multicolumn{6}{l}{$^*$ Gammatone filter bank. \# Trainable parameters: 1,223,082}
		\end{tabular}
\end{table}

Koerich~\etal\cite{Koerich2020} proposed a 1D CNN architecture for music genre classification to demonstrate the transferability of adversarial audio attacks from spectrograms to audio waveforms. Such an architecture is based on the previous architecture proposed by Abdoli~\etal\cite{abdoli2019}, but it is adapted to music audio. It also uses a CL initialized as a Gammatone filter bank and several CLs with large filters. Average pooling layers (APLs) replace the first two MPLs, and LeakReLU replaces the ReLU activation functions at all CLs. A single FCL is used before the output layer with a dropout with a probability of 0.4. Finally, batch normalization is used after each CL. Table~\ref{tab:AbdoliAdapCNN} shows such a 1D CNN architecture.

\begin{table}[htpb!]
\renewcommand{\arraystretch}{1.1}
\caption{CNN architecture proposed by Koerich~\etal\cite{Koerich2020} adapted from~\cite{abdoli2019}.}
\label{tab:AbdoliAdapCNN}
	\centering
		\begin{tabular}{|c|c|c|c|c|c|c|c|}
			\hline
			\bf Layer & \bf \# Filters & \bf Kernel & \bf Pool & \bf Stride & \bf Output \\
			 &  &  & \bf Size &  & \bf Shape\\
			\hline
			Input & - & - & - & - & 110,250\\
			\hline
			Conv1D$^*$ & 32 & 512 & - & 1 & 32$\times$109,739\\
			\hline
			AvgPool & - & - & 8 & 8 & 32$\times$13,717\\
            \hline
            Conv1D & 16 & 256 & - & 2 & 16$\times$6,731\\
            \hline
			AvgPool & - & - & 8 & 8 & 16$\times$841\\
            \hline
            Conv1D & 32 & 64 & - & 2 & 32$\times$389\\
            \hline
            Conv1D & 64 & 32 & - & 2 & 64$\times$179\\
            \hline
            Conv1D & 128 & 16 & - & 2 & 128$\times$82\\
            \hline
			MaxPool & - & - & 2 & 2 & 128$\times$41\\
            \hline
			FC & - & - & - & - &256\\
            \hline
            Output & - & - & - & - & 10 \\
            \hline
            \multicolumn{6}{l}{$^*$ Gammatone filter bank. \# Trainable parameters: 1,707,506}
		\end{tabular}
\end{table}

\section{EXPERIMENTAL RESULTS}
\label{sec:exp}
This section assesses the performance of the proposed residual 1D CNN presented in Section~\ref{sec:propo} and compares its performance with other 1D CNNs described in Section~\ref{sec:1DCNNs}.

\subsection{Dataset}
The 1D CNN architectures for music genre classification were evaluated on the GTzan dataset. This dataset consists of 1,000 30-second music excerpts evenly distributed into ten classes: rock, reggae, blues, classical, disco, country, hip-hop, metal, jazz, and pop. The audio samples were collected from a variety of sources to represent a variety of recording conditions. The GTzan dataset has several known integrity problems, such as replications, mislabeling, and distortions.

The experimental protocol employs a stratified 3-fold holdout where the 1,000 audio files were shuffled and divided into three folds with 340, 330, and 330 samples. Fold 1 contains 34 music tracks of each of the ten genres; Folds 2 and 3 have 33 music tracks of each genre, respectively. Next,  every music track of every fold was split into 21 short segments (5 sec) using a sliding window according to the approach shown in Fig.~\ref{fig:framing}.

\subsection{Experimental Protocol and Results}
In the first round, Fold 1 is used for training, Fold 2 for validation, and Fold 3 for testing. Next, we rotate the folders, and at the end, each partition is used for training, validation, and test. Table~\ref{tab:perfNoDA} shows the results of the first experiment that compares the classification accuracy and standard deviation of each 1D CNN averaged over the three rounds. Each 1D CNN was trained up to 100 epochs with batch sizes of 80 and early stopping. After predicting the music genre for each segment on the test set, the predictions for all 21 audio segments belonging to the same song are aggregated either with a majority vote or a sum rule to determine the final genre prediction for the whole music track.

\begin{table}[htpb!]
\renewcommand{\arraystretch}{1.1}
\caption{Average accuracy and standard deviation achieved by the 1D CNN architectures on 3-fold of GTzan dataset. Results without data augmentation.}
\label{tab:perfNoDA}
	\centering
		\begin{tabular}{|r|c|c|c|}
			\hline
			&  & \bf Comb &  \\
			\multicolumn{1}{|c|}{\bf Reference} & \bf Segments & \bf Rule & \bf Aggregation \\
			\hline
			Choi~\etal\cite{Choi2017} & 69.28$\pm$0.76 & Maj & 75.99$\pm$0.56 \\
            \hline
            Pons~\etal\cite{Pons2018} & 63.41$\pm$0.76 & Maj & 70.51$\pm$1.36 \\
            \hline
			Dieleman \& Schrauwen~\cite{Dieleman2014} & 45.52$\pm$1.92 & Maj & 48.50$\pm$1.62 \\
            \hline
			Abdoli~\etal\cite{abdoli2019} & 14.09$\pm$3.56 & Maj & 14.96$\pm$3.77 \\
            \hline
			Koerich~\etal\cite{Koerich2020} & 63.19$\pm$1.79 & Sum & 69.12$\pm$2.12 \\
            \hline
            1D ResNet CNN & 72.88$\pm$1.75 & Sum & 76.02$\pm$1.60 \\
            \hline
            \multicolumn{4}{l}{}
		\end{tabular}
\end{table}

\subsection{Data Augmentation}
Since the amount of data for training 1D CNNs play an important role in the final accuracy of such architectures, we used five signal transformations to generate artificial audio samples. First, Gaussian noise adds to the audio signal a Gaussian noise of random amplitude between 0.005 and 0.02. Second, gain multiplies the audio signal by a random amplitude factor between -12 dB and 12 dB to reduce or increase the audio signal's volume. Third, loudness normalization applies a constant amount of gain to match a specific loudness according to ITU-R BS.1770-4~\cite{ITU}. Fourth, pitch shifting shifts the sound up and down between 8 and -8 semitones without changing the tempo. Finally, time stretches the signal with a random rate between 0.5 and 1.5 without changing the pitch. These data augmentation operations increase the size of the original dataset by the factor of five since they are applied individually, and they generate five extra audio samples for every original audio sample.

Table~\ref{tab:perfDA} shows the results of the second experiment that compares the classification accuracy of each 1D CNN trained with data augmentation.

\begin{table}[htpb!]
\renewcommand{\arraystretch}{1.1}
\caption{Average accuracy and standard deviation achieved by the 1D CNN architectures on 3-fold of GTzan dataset. Results with 5$\times$ data augmentation.}
\label{tab:perfDA}
	\centering
		\begin{tabular}{|r|c|c|c|}
			\hline
			&  & \bf Comb &  \\
			\multicolumn{1}{|c|}{\bf Reference} & \bf Segments & \bf Rule & \bf Aggregation \\
			\hline
			Choi~\etal\cite{Choi2017} & 71.23$\pm$1.03 & Sum & 79.32$\pm$1.83 \\
            \hline
            Pons~\etal\cite{Pons2018} & 66.31$\pm$1.35 & Maj & 74.32$\pm$1.64 \\
            \hline
			Dieleman \& Schrauwen~\cite{Dieleman2014} & 45.20$\pm$2.22 & Sum & 53.62$\pm$2.40 \\
            \hline
			Abdoli~\etal\cite{abdoli2019} & 15.00$\pm$3.86 & Maj & 15.59$\pm$4.17 \\
            \hline
			Koerich~\etal\cite{Koerich2020} & 64.90$\pm$1.73 & Sum & 73.02$\pm$1.10 \\
            \hline
			1D ResNet CNN & 74.62$\pm$1.91 & Sum & 80.93$\pm$2.35 \\
            \hline
            \multicolumn{4}{l}{}
		\end{tabular}
\end{table}

\subsection{Discussion}
The best results reported in Tables~\ref{tab:perfNoDA} and~\ref{tab:perfDA} were achieved by architectures that use small convolutional filters at the first layers (size 3), namely 1D ResNet CNN,~\cite{Choi2017}, and ~\cite{Pons2018}. On the other hand, the architecture that employed large convolutional filters (size 512) achieved the worst average accuracy~\cite{abdoli2019}. This corroborates with the findings of Kim~\etal\cite{Kim2018} that filters with small granularity in time for all Cls perform better than large ones. 

Data augmentation improves the performance of all 1D CNN architectures, reducing the overfitting and improving the generalization of all models. However, the downside is that the time for training the 1D CNNs increases almost linearly with the number of samples.

A fair comparison with other shallow or deep approaches is challenging due to the differences in the experimental protocols. Several shallow approaches use 10-fold cross-validation, which is not feasible when evaluating CNN architectures due to the limitation of computational resources. Kim~\etal\cite{Kim2018} compared the performance of several approaches for music genre classification on the GTzan dataset. The accuracy reported by Kim~\etal\cite{Kim2018} ranges between 63.20\% and 82.10\%, where the best result was achieved using transfer learning with a 1D CNN pre-trained on 189,189 songs.    

\section{CONCLUSIONS}
This paper proposed a 1D residual CNN architecture for music genre classification. The network architecture consists of two CLs and nine residual blocks, and it learns the representation directly from segments of the audio signal. The proposed 1D residual CNN was evaluated on a dataset of 1,000 excerpts of music pieces, and the experimental results have shown that the proposed architecture outperforms other 1D CNN architectures. Furthermore, the proposed 1D residual CNN performs relatively well compared to the state-of-the-art on the GTzan dataset.

The amount of data available for training all 1D architectures plays an important role in the final accuracy. The experimental results have shown an average increase of 3.62\% in the accuracy achieved by the 1D CNNs. Therefore, we will pre-train the proposed 1D residual CNN on large music datasets such as the Latin music dataset LMD~\cite{silla2008latin}, Free music archive dataset~\cite{Defferrard2017}, and MSD dataset~\cite{Kim2018}, as future work. We believe that we can achieve even better results.


\balance
\bibliographystyle{./bibliography/IEEEtran}

\end{document}